\documentclass[manuscript]{acmart}

\AtBeginDocument{%
  \providecommand\BibTeX{{%
    \normalfont B\kern-0.5em{\scshape i\kern-0.25em b}\kern-0.8em\TeX}}}

\begin{document}

\title{AI Literacy in K-12 and Higher Education in the Wake of Generative AI: An Integrative Review}


\author{Xingjian (Lance) Gu}
\email{xjgu@umich.edu}
\orcid{0009-0000-0433-9843}
\affiliation{%
  \institution{University of Michigan School of Information}
  \country{United States}
}

\author{Barbara J. Ericson}
\orcid{0000-0001-6881-8341}
\email{barbarer@umich.edu}
\affiliation{%
  \institution{University of Michigan School of Information}
  \country{United States}
}


\begin{abstract}
Even though AI literacy has emerged as a prominent education topic in the wake of generative AI, its definition remains vague. There is little consensus among researchers and practitioners on how to discuss and design AI literacy interventions. 
The term has been used to describe both learning activities that train undergraduate students to use ChatGPT effectively and having kindergarten children interact with social robots. This paper applies an integrative review method to examine empirical and theoretical AI literacy studies published since 2020, to identify shifting definitions and emerging trends in AI literacy around the public introduction of generative AI. In synthesizing the 124 reviewed studies, three ways to conceptualize literacy---functional, critical, and indirectly beneficial---and three perspectives on AI---technical detail, tool, and sociocultural---were identified, forming a framework that reflects the spectrum of how AI literacy is approached in practice. The framework highlights the need for more specialized terms within AI literacy discourse and indicates research gaps in certain AI literacy objectives.
\end{abstract}



\maketitle

\section{Introduction}

The rapid development of AI-based technologies brings about promising potentials and risks. Accordingly, education researchers and practitioners have increasingly turned to AI literacy as an important learning objective. 
However, the definition of AI literacy remains vague. Researchers have used the term to describe learning interventions that differ by in school contexts, learning objectives, and types of AI technologies they use. Furthermore, the research of AI literacy is shifting significantly in the wake of generative AI. Thus, it is crucial to review the field and develop a conceptual framework that captures the diverse conceptualizations of AI literacy.

The concept of AI literacy and recognition of its potential significance are well-established \citep{touretzky_envisioning_2019, long_what_2020}. One of the pioneering works by Touretzky et al. in 2019 laid out “five big ideas” for the AI4K12 initiative: “computers perceive the world using sensors”, “agents maintain models/representations of the world and use them for reasoning”, “computers can learn from data”, “making agents interact with humans is a substantial challenge for AI developers”, and “AI applications can impact society in both positive and negative ways” \citep{touretzky_envisioning_2019}. This paper had a major influence on subsequent AI literacy curriculum design. The next year, another prominent work by Long and Magerko defined AI literacy as “a set of competencies that enables individuals to critically evaluate AI technologies; communicate and collaborate effectively with AI; and use AI as a tool online, at home, and in the workplace” \citep{long_what_2020}. However, even among these early definitions, there already were some differences, such as whether AI literacy is more about understanding the technical foundations of AI, the use of AI tools, or about critical evaluation of AI.

AI has always been loosely defined, even as a technical concept. Originally denoting search algorithms in the 50s, AI shifted to mean expert systems in the 80s, machine learning models in the 2000s, and now large language models in the early 2020s \citep{haenlein_brief_2019}. Beyond its technical definitions, AI is applied in many technologies across different contexts. Thus, it cannot be understood merely as a technical discipline, but as sociotechnical systems \citep{winner_artifacts_1985}. When evaluated in this way, the use of AI has raised ethical concerns, including AI accountability for its decisions, latent biases, environmental costs, and privacy intrusions during training data collection \citep{awad_moral_2018, buolamwini_gender_2018, selinger_ethics_2021}. Crucially, these sociocultural effects of AI are connected to, but distinct from understanding AI's underlying technical mechanisms. 

Literacy, while originally referring to the ability to read and write, has been adopted to describe various technology-related skill sets, including digital literacy, computational literacy, data literacy, among others \citep{eshet_digital_2004, gummer_building_2015, disessa_computational_2018}. 
In practice, “literacy” could entail diverse learning objectives. For instance, Selber theorized that digital literacy is comprised of functional literacy, critical literacy, and rhetorical literacy (the ability to self-express using digital technologies), together forming digital multiliteracies \citep{selber_multiliteracies_2004}. Extending these ideas to AI technologies, researchers have considered both the skills of using AI technologies and understanding their broader sociocultural implications as literacy. 

Further complicating the concept of AI literacy, the rise of publicly available generative AI tools rapidly changed the landscape of AI. The vast increase in ability of LLMs to perform a wide range of tasks, coupled with lowered barriers of access given their natural language interface, prompted researchers to rethink what future students could do with the help of AI \citep{denny_conversing_2023, finnie-ansley_my_2023, phung_generative_2023, prather_robots_2023, joshi_chatgpt_2024, padiyath_insights_2024}. Amid reflections on how generative AI tools could reshape computing education objectives \citep{kamoun_exploring_2024, lau_ban_2023, rahman_chatgpt_2023}, teaching AI literacy emerged as a potential path forward. The resulting proliferation of AI literacy research brought greater diversity to the field, resulting in a wide range of definitions of AI literacy.

AI literacy is thus an umbrella term that can encompass very dissimilar interpretations that are equally valid. Educators and scholars have labeled initiatives that have completely different pedagogical approaches and objectives as “AI literacy”, ranging from having kindergarteners interact with social robots to helping CS major undergraduate students practice prompt engineering for LLMs \citep{su_span_2024, denny_prompt_2024}. It is thus necessary to create a framework to organize the work, in order to create more precise terms and facilitate more effective and clear communication among researchers and practitioners. To achieve that, this study is guided by the following research questions:

\textbf{\textit{RQ1: How do researchers conceptualize and approach AI literacy in K-12 and higher education?}}

\textbf{\textit{RQ2: What new trends have emerged in AI literacy research in K-12 and higher education in the wake of generative AI?}}

This study conducts an integrative review of 124 studies published between 2020 and 2024. By analyzing what forms of AI the studies address and what capabilities they seek to promote in students, this review derives common conceptualizations of AI literacy and presents them under a unified framework. 

This study contributes to existing AI literacy work in three ways. 1) It updates prior AI literacy reviews and definitions by including new AI literacy studies published since the introduction of generative AI, 2) it delineates the boundaries of AI literacy to highlight the need for more precise and informative terms, and 3) it adapts established theories in digital literacy to build a unifying framework for AI literacy. This framework does not seek to supplant prior lists of competencies, but situates them in a broader theoretical landscape, so they can be in conversation with each other. By capturing the current AI literacy discourse in a consistent framework, this paper seeks to promote more effective communication among researchers and practitioners by advocating for terms that can better describe their objectives and approaches than generic “AI literacy”.

\section{Prior Work}
\subsection{AI literacy Frameworks and Reviews}
Long and Magerko created the most influential framework by describing a set of seventeen competencies related to AI \citep{long_what_2020}. Their definition of AI literacy as “a set of competencies that enables individuals to critically evaluate AI technologies; communicate and collaborate effectively with AI; and use AI as a tool online, at home, and in the workplace” is widely cited in empirical work on AI literacy interventions \cite{long_what_2020}. A group of University of Hong Kong researchers also conducted a series of highly cited AI literacy reviews. Their work included both high-level overviews of the conceptualization of AI literacy \citep{ng_ai_2021, ng_conceptualizing_2021} and reviews of empirical studies within various contexts, including K-12 education in the Asia-Pacific region \citep{su_meta-review_2022}, early childhood \citep{su_artificial_2023-1}, and secondary education \citep{ng_fostering_2024-1}. They proposed four aspects of AI literacy, “know and understand, use and apply, evaluate and create, and ethical issues”, and define AI literacy as a series of competencies similar to those of Long and Magerko \citep{long_what_2020, ng_conceptualizing_2021}. Since these studies were published prior to the mainstream introduction of generative AI, they may need an update. 

Several reviews explicitly acknowledged the nascent and changing nature of AI literacy and the resulting difficulty in defining it \citep{laupichler_artificial_2022, benvenuti_artificial_2023, tenorio_artificial_2023, lee_systematic_2024}. Other reviews have proposed frameworks to capture the multitudes of AI literacy, including a twelve-faceted taxonomy synthesized from prominent AI concept classifications \citep{shiri_artificial_2024}, a competency-based framework similar to the work by Ng et al. \citep{almatrafi_systematic_2024}, and two learner-oriented frameworks that categorize AI literacy interventions based on learners’ roles in relation to AI \citep{faruqe_competency_2022, schuller_better_2023}. While these taxonomies are comprehensive, they do not address educators' radically different interpretations of AI literacy and the resulting implementations. This review's proposed framework focuses on the conceptualizations of AI literacy, thus yielding insights into the motivations behind different AI literacy approaches.


\subsection{Definitions of AI and Literacy}
Complicating the matter of defining AI literacy, both constituent concepts of AI and literacy are complex and lack consensus definitions. To start with AI, the term has shifted to encompass different technologies and paradigms since its conception in the 1950s \citep{haenlein_brief_2019}. In its technical sense, AI usually refers to machine learning and generative AI technologies in the reviewed studies. Older paradigms in AI, such as knowledge representation and reasoning, are almost never mentioned among the studies included in this review. 

However, it is insufficient to understand AI only as a technical field. AI's has been integrated into society to form sociotechnical systems, with social, cultural, and political implications \citep{winner_artifacts_1985}. To capture the full breadth of AI conceptualizations, this review draws from the Dagstuhl Triangle, which was originally developed in Germany to describe the objectives of teaching learners to interact with digital systems in general \citep{brinda_dagstuhl-erklarung_2016}. It involves three perspectives: the technological perspective of understanding how the digital system functions, the user-oriented perspective of how to use the digital system, and the sociocultural perspective of how the digital system interacts with society \citep{michaeli_data_2023}. Similar theoretical frameworks have been applied to AI literacy. For instance, Kong et al. expressed their AI literacy learning objectives in terms of “cognitive”, “affective”, and “sociocultural” \citep{kong_evaluating_2023}. The Dagstuhl Triangle scaffolds the conceptual framework of this review, as the perspectives effectively capture the prominent ways AI literacy scholars discuss AI. 

The concept of literacy is similarly difficult to define. “Literacy” is borrowed from its original context of reading and writing, which adds some confusion. Even in its original meaning, literacy is complex, entailing different objectives and approaches \citep{scribner_literacy_1984}. In her 1984 work, Scribner described literacy in three metaphors, as adaptation---the functional knowledge to perform reading and writing tasks as society requires, as power---being empowered to communicate and organize for community advancement, and as state of grace---self-improvement that signals greater social status, virtue, or intelligence \citep{scribner_literacy_1984}. Scribner’s underlying perspectives on literacy---as functional knowledge, as empowerment and critical thinking, and as indirectly beneficial virtues---are valuable when making sense of the divergent motivations behind promoting AI literacy.

More closely related to AI literacy, there is an extensive body of literature on digital literacy and other technological literacies, such as data literacy \citep{gummer_building_2015} and computational literacy \citep{disessa_computational_2018}. Echoing Scribner’s three metaphors of literacy, Selber described an influential framework of digital multiliteracies, which includes functional, critical, and rhetorical literacies \citep{selber_multiliteracies_2004}. Selber's functional literacy and critical literacy share significant similarities with Scribner’s first two metaphors, meaning functional skills of using digital technologies and critical evaluation of digital artifacts respectively \citep{selber_multiliteracies_2004}. Selber’s rhetorical literacy is a combination of functional and critical capabilities that enable individuals to create new digital media to express themselves \citep{selber_multiliteracies_2004}. Other digital literacy frameworks are more focused on listing fine-grained competencies \citep{eshet_digital_2004}, which reflects some of the aforementioned AI literacy reviews.

\section{Integrative Review Methodology}
\subsection{Rationale}
Since 2019, there has been a significant increase in the number of publications on AI literacy \citep{lee_systematic_2024, tenorio_artificial_2023}. However, the authors of these studies frequently acknowledge that AI literacy is not clearly defined. They typically adopt Long and Magerko's definition despite seismic shifts in AI since then \citep{almatrafi_systematic_2024, lee_systematic_2024, long_what_2020}. 

The lack of a clear definition and major developments in AI technologies pose several challenges to conducting a literature review on AI literacy. Even though AI literacy is traditionally of interest to AI education researchers, the newfound popularity of Large Language Model (LLM) based tools has attracted research interest from other fields. These include human-computer interaction (HCI), computer science education, and learning sciences. There are also a significant number of both intervention studies and theoretical work aimed at defining AI literacy, as opposed to mostly intervention studies. Lastly, AI literacy intervention studies can have completely different learning objectives, underlying mechanisms, methodologies and study designs, even when controlling for student populations and school settings. 

These factors render some review methodologies unsuitable, such as meta-analysis, which are used to evaluate the effectiveness of specific interventions given their empirical evidence, or systematic review, which is used to adjudicate on specific research questions\citep{souza_integrative_2010, cronin_why_2023}. In contrast, the integrative review method seeks to integrate both theoretical and empirical literature to generate new comprehensive insights on a topic, as opposed to adjudicating established positions \citep{whittemore_integrative_2005, cronin_why_2023}. Integrative reviews build structured conceptual frameworks that capture holistic patterns across different communities of practice, instead of delving into individual findings \citep{cronin_why_2023}. The research question of this study explores overarching trends, rather than specific learning interventions or outcomes. Since AI literacy as a research field has vaguely defined boundaries, includes a large number of both theoretical and empirical works, and is studied by researchers across several fields, the integrative review method is the most suitable to review studies in the field.

An integrative review ensures rigor by identifying the connected but distinct discourse about a topic, constructing a conceptual structure to organize them, and generating new knowledge from this integration \citep{torraco_writing_2005, cronin_why_2023}. Thus, the integrative review methodology is particularly suitable for addressing emerging topics and building new frameworks \citep{torraco_writing_2005}. This methodology has often been applied in nursing science, mental health research, and organization studies \citep{torraco_writing_2005}, and it can be used to manage the complexity and diversity of AI literacy studies. 

\subsection{Methodology}
\subsubsection{Scoping and Problem Identification}
Following guidance on conducting integrative reviews \citep{whittemore_integrative_2005, cronin_why_2023}, the process begins with determining the purpose of the review and identifying the problem to be addressed. A purposive search was first conducted, which entails searching for studies to address a question without concerning about the balance or comprehensiveness of the search. To identify prominent recent AI literacy studies, the purposive search focused on studies from high-visibility venues, such as ICER, ITiSCE, and SIGCSE. This search revealed the wide range of conceptualizations of AI literacy and corresponding diverse interventions, from technical machine learning courses to critical evaluation of AI applications, and thus motivated the creation of a conceptual framework.

\subsubsection{Literature search}
In order to fully integrate literature across different methodologies and fields, the literature search process of an integrative review aims to minimize bias towards any given field and seeks to be complete and balanced \citep{cronin_why_2023}. As such, the literature search stage applies a systematic and exhaustive search approach. To ensure balanced coverage of the fields relevant to AI literacy, two digital database were used---the Education Resources Information Center (ERIC) and Scopus, which includes literature from various venues affiliated with Association for Computing Machinery (ACM). The search included various terms associated with AI and literacy, as shown in table \ref{search_term}. To complement the results of searching digital databases, we conducted a second round of purposive search by tracing citations of the included studies. 
The search included sources published between 2020 and July 2024. The end point of July 2024 was when the search was conducted. The starting point of 2020 was set to include AI literacy studies from before the public introduction of generative AI, so changes in definitions and new trends could be observed. The starting point was not set earlier to ensure the AI technologies discussed by the studies were relatively contemporary and comparable to each other. The inclusion and exclusion criteria are described in table \ref{ICEC}.



\begin{table}[ht]
\small
\begin{tabular}{p{0.33\linewidth}p{0.6\linewidth}}
\hline
Category                  & Search Keywords                                                                                   \\ \hline
Terms related to AI       & machine learning OR ML OR artificial intelligence OR AI OR large language model OR LLM OR ChatGPT \\ \hline
Terms related to Literacy & literacy OR competency                                                                            \\ \hline
Logistical terms          & published since 2020 AND in English AND in peer-reviewed publications                             \\ \hline
\end{tabular}
\caption{The search terms used in the digital databases}
\label{search_term}
\end{table}

\begin{table}[ht]
\small
\begin{tabular}{p{0.08\linewidth}p{0.84\linewidth}}
\hline
Criteria                  & Criteria Details                                                                                   \\ \hline
IC1       & The study should be a peer reviewed publication in journals or conference proceedings. \\ \hline
IC2 & The study can be theoretical or empirical, including curriculum design, intervention study, developing AI literacy survey instruments, interviewing experts about their AI literacy definition, etc. This is to achieve the flexibility necessary for an integrative review.                                                                         \\ \hline
IC3          & The study should be published in English.                           \\ \hline
IC4          & The study should be published since 2020.                             \\ \hline
IC5          & The study clearly communicates its objectives and motivations for studying AI literacy, since this review focuses on conceptualizations rather than interventions.                           \\ \hline
IC6          & The education setting of the study can be across the whole world. AI literacy is garnering interest and active research efforts from around the world, and AI literacy research can have diverse standpoints and approaches by different researchers. Thus, this criteria is to ensure balanced and just representation of their perspectives.                            \\ \hline \hline
EC1          & The education stage should be K-12 or undergraduate, thus excluding graduate schools, adult upskilling and reskilling, professional training, etc. Notably, teacher AI literacy is a popular topic among recently published studies and is relevant to the broader discussion of promoting AI literacy among students, but was deemed out of scope for this work.                            \\ \hline
EC2          & The study should not involve discipline-specific AI literacy other than Computer Science. This excludes AI literacy specific to healthcare, business, English as a second language (ESL), etc. This criteria does not exclude general AI literacy, such as teaching AI literacy to non-CS major students.                          \\ \hline
EC3          & The study should not be about AI-based learning, such as theorizing applications of LLM in education or building and testing AI-based tools. In other words, included literature should be about learning about AI, not learning with AI. As such, educational tools developed to promote AI literacy should be included.                             \\ \hline
EC4          & The study should not be only measuring or observing students’ perception of AI, such as surveying student AI-usage behavior or understanding of AI. The study should involve teaching or promoting AI-related skills or awareness, or deriving AI literacy outcomes by consulting with students.                           \\ \hline
\end{tabular}
\caption{The inclusion and exclusion criteria used when assessing sources}
\label{ICEC}
\end{table}

\subsubsection{Data Evaluation}
\begin{figure}[ht]
    \centering
    \includegraphics[width=0.6\linewidth]{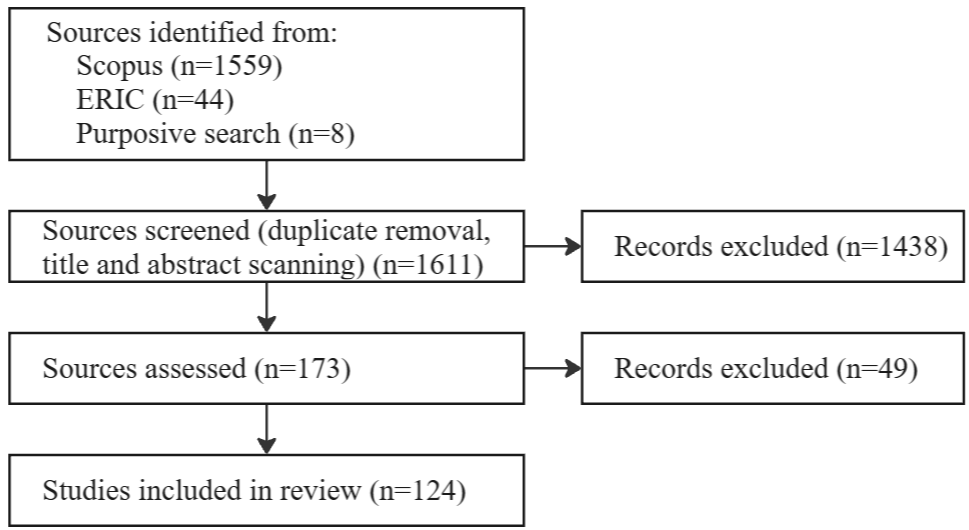} \\
    \caption{Literature search process and number of studies after each step}
    \Description{A flowchart that shows the literature search process and number of studies after each step.}
    \label{lit_flowchart}
    \vspace{-3mm}
\end{figure}
The literature search process is shown in figure \ref{lit_flowchart}. The initial search term on the digital databases yielded 1559 results on Scopus and 44 on ERIC. After removing duplicate results and reviewing the study titles and abstracts for relevance to the research topic, there remained 153 results from Scopus and 12 from ERIC. Next, studies were evaluated using the inclusion and exclusion criteria to yield 107 sources from Scopus and 9 from ERIC. 
Eight studies were added from the purposive search. One of them, Touretzky et al.’s Five Big Ideas in AI, was published in 2019, but is highly influential in AI literacy discourse and thus included \citep{touretzky_envisioning_2019}.

\subsubsection{Data Extraction and Analysis}
To address the research questions, the data extraction process focused on each included study's statements regarding their conceptualization of AI literacy. These included their definition of the term, the studies they cited, the motivations and goals behind researching AI literacy and developing their intervention, and their research questions. The goal of data extraction and analysis of an integrative review is a thematic synthesis to reach a higher level of abstraction \citep{cronin_why_2023}. To that end, the process began with a bottom-up approach by coding the different themes of how AI and literacy are described and iteratively refining the codes to build a more concise conceptual framework. For instance, some earlier AI conceptualization codes include ‘machine learning’, ‘non-ML technical knowledge’, ‘LLM tools’, and ‘technological applications’. After solidifying the conceptual structure and identifying the theoretical framework of the Dagstuhl Triangle \citep{brinda_dagstuhl-erklarung_2016}, the data analysis shifted toward a more top-down approach. The theoretical frameworks were used to update the codes to better support the abstract conceptual structure \citep{cronin_why_2023}. 

\section{Results}

\subsection{Conceptual Framework}

\begin{figure}[ht]
    \centering
    \includegraphics[width=0.8\linewidth]{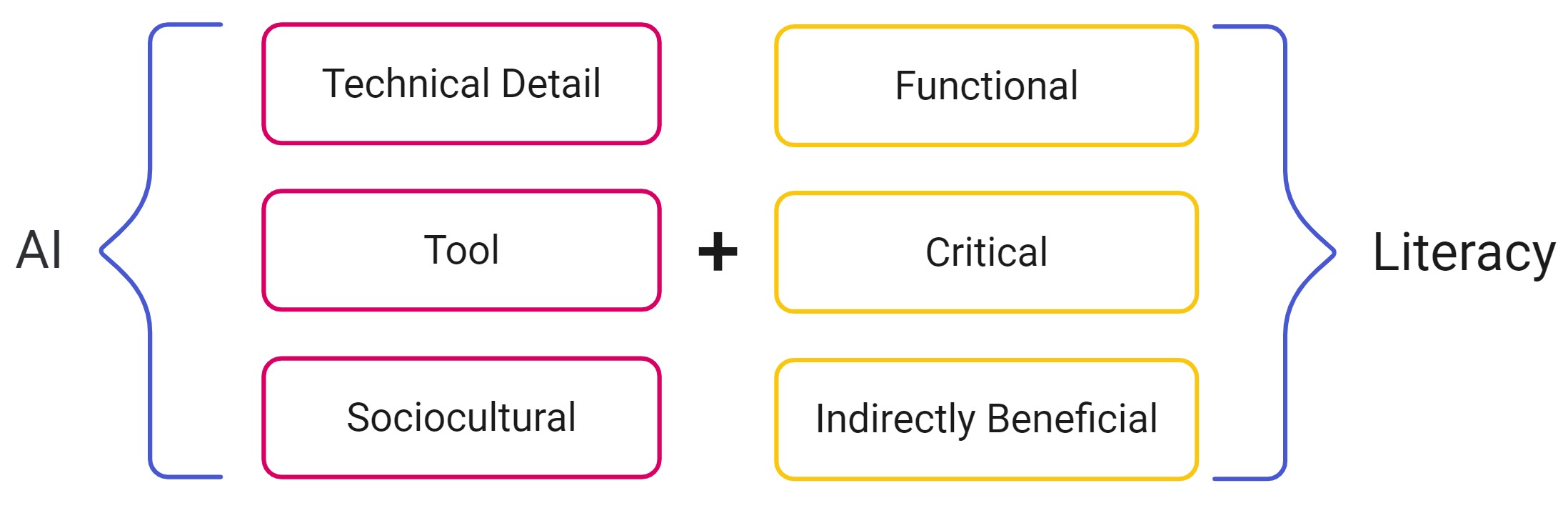}
    \caption{Overview of the Conceptual Framework Based on the Perspectives of AI and Perspectives of Literacy}
    \Description {Overview of the Conceptual Framework Based on the Perspectives of AI and Perspectives of Literacy}
    \label{framework}
\end{figure}

This review applied the conceptual framework in figure \ref{framework}. The framework captures the diversity in AI literacy definitions with three conceptualizations of AI and three conceptualizations of literacy, which can be freely combined and are not mutually exclusive. The conceptualizations are derived from the reviewed studies and informed by theories including the Dagstuhl Triangle and digital multiliteracies \citep{brinda_dagstuhl-erklarung_2016, selber_multiliteracies_2004}.

The first conceptualization of AI is AI as technical details. An example AI literacy intervention that adopts a technical detail perspective would be teaching supervised learning algorithms. The second is AI as tools, focusing on tools using AI without engaging with their inner workings. An example of a tool perspective AI literacy intervention would be teaching effective use of ChatGPT. The third conceptualization is AI sociocultural impact. An example would be teaching the gender and racial biases in facial recognition technology.

The first conceptualization of literacy is functional literacy. A learning goal with a functional literacy perspective would be to prepare learners for jobs that would use AI. The second is critical literacy. An example goal would be to prepare learners to become informed citizens who can weigh in on AI regulation policies. Lastly, an example goal from the indirectly beneficial literacy perspective would be making learners become more interested in STEM in general. 

The conceptualizations are combined to describe AI literacy interventions. For instance, a curriculum that teaches both machine learning fundamentals and how they can cause algorithmic biases in applications, which seeks to promote both proficiency and critical awareness of AI technologies, would have both technical details and sociocultural perspectives of AI, and both functional and critical perspectives of literacy.



\subsection{Overview of Reviewed Studies}

\begin{table}[htbp]
\small
\begin{tabular}{p{0.25\linewidth}p{0.45\linewidth}p{0.05\linewidth}p{0.07\linewidth}p{0.06\linewidth}}
\hline
Author                                                        & Title                                                                                                                                                              & Year                         & Venue   & Citation \\ \hline
Touretzky, D., Gardner-McCune, C., Martin, F., \& Seehorn, D. & Envisioning AI for K-12: What Should Every Child Know about AI?                                                                                                    & 2019                         & AAAI    & 615       \\ \hline
Long D.; Magerko B.                                           & What is AI Literacy? Competencies and Design Considerations                                                                                                        & 2020                         & CHI     & 512       \\\hline
Ng D.T.K.; Leung J.K.L.; Chu S.K.W.; Qiao M.S.                & Conceptualizing AI literacy: An exploratory review                                                                                                                 & 2021                         & CAEAI & 239       \\\hline
Zamfirescu-Pereira J.D.; Wong R.Y.;   Hartmann B.; Yang Q.    & Why Johnny Can't Prompt: How Non-AI Experts Try (and Fail) to Design LLM Prompts                                                                                   & 2023                         & CHI     & 157       \\\hline
Lee I.; Ali S.; Zhang H.; Dipaola D.; Breazeal C.             & Developing Middle School Students' AI Literacy                                                                                                                     & 2021                         & SIGCSE  & 112       \\\hline
Su J.; Zhong Y.; Ng D.T.K.                                    & A meta-review of literature on educational approaches for teaching AI at the K-12 levels in the Asia-Pacific   region                                              & 2022                         & CAEAI & 82        \\\hline
Laupichler M.C.; Aster A.; Schirch J.; Raupach T.             & Artificial intelligence literacy in higher and adult education: A scoping literature review                                                                        & 2022                         & CAEAI & 80        \\\hline
Ng D.T.K.; Leung J.K.L.; Chu K.W.S.; Qiao M.S.                & AI Literacy: Definition, Teaching, Evaluation and Ethical Issues                                                                                                   & 2021                         & ASIS\&T & 78        \\\hline
Kong, S. C., Cheung, W. M. Y., \& Zhang, G.                   & Evaluating an Artificial Intelligence Literacy Programme for Developing University Students' Conceptual Understanding, Literacy, Empowerment and Ethical Awareness & 2023                         & ET\&S   & 67        \\\hline
Su J.; Ng D.T.K.; Chu S.K.W.                                  & Artificial Intelligence (AI) Literacy in Early Childhood Education: The Challenges and Opportunities                                                               & 2023 & CAEAI & 66        \\ \hline
\end{tabular}
\caption{The Ten Reviewed Studies with the Most Citations}
\label{top_ten}
\end{table}

Table \ref{top_ten} lists the ten most cited studies among the reviewed studies. The two most cited studies by Touretzky et al. and Long \& Magerko laid the foundations for defining AI literacy by listing skills and competencies that they deemed central to being AI literate \citep{touretzky_envisioning_2019, long_what_2020}. Five of the studies are literature reviews, covering AI literacy in early childhood, K-12, post-secondary, and adult education contexts \citep{ng_conceptualizing_2021, su_meta-review_2022, laupichler_artificial_2022, ng_ai_2021, su_artificial_2023-1}. However, all of these reviews were conducted before the rise of generative AI, indicating a need for new reviews to update our understanding of the AI literacy research landscape. Three studies are empirical, discussing "LLM-and-prompt literacy” \citep{zamfirescu-pereira_why_2023}, middle school AI literacy curriculum \citep{lee_developing_2021}, and teaching machine learning to college non-CS major students \citep{kong_evaluating_2023}. The diversity of subject matter in these empirical studies demonstrates the complexity of AI literacy as a concept and highlights the need for a conceptual framework to organize them. The venues where these studies were published are similarly diverse, with both specialized education venues, such as SIGCSE and Computer \& Education: Artificial Intelligence (CAEAI), and more general venues, such as CHI and Education Technology \& Society (ET\&S).

\begin{figure}[!htbp]
  \centering
  \hfill
  \begin{minipage}[b]{0.4\textwidth}
    \includegraphics[width=\textwidth]{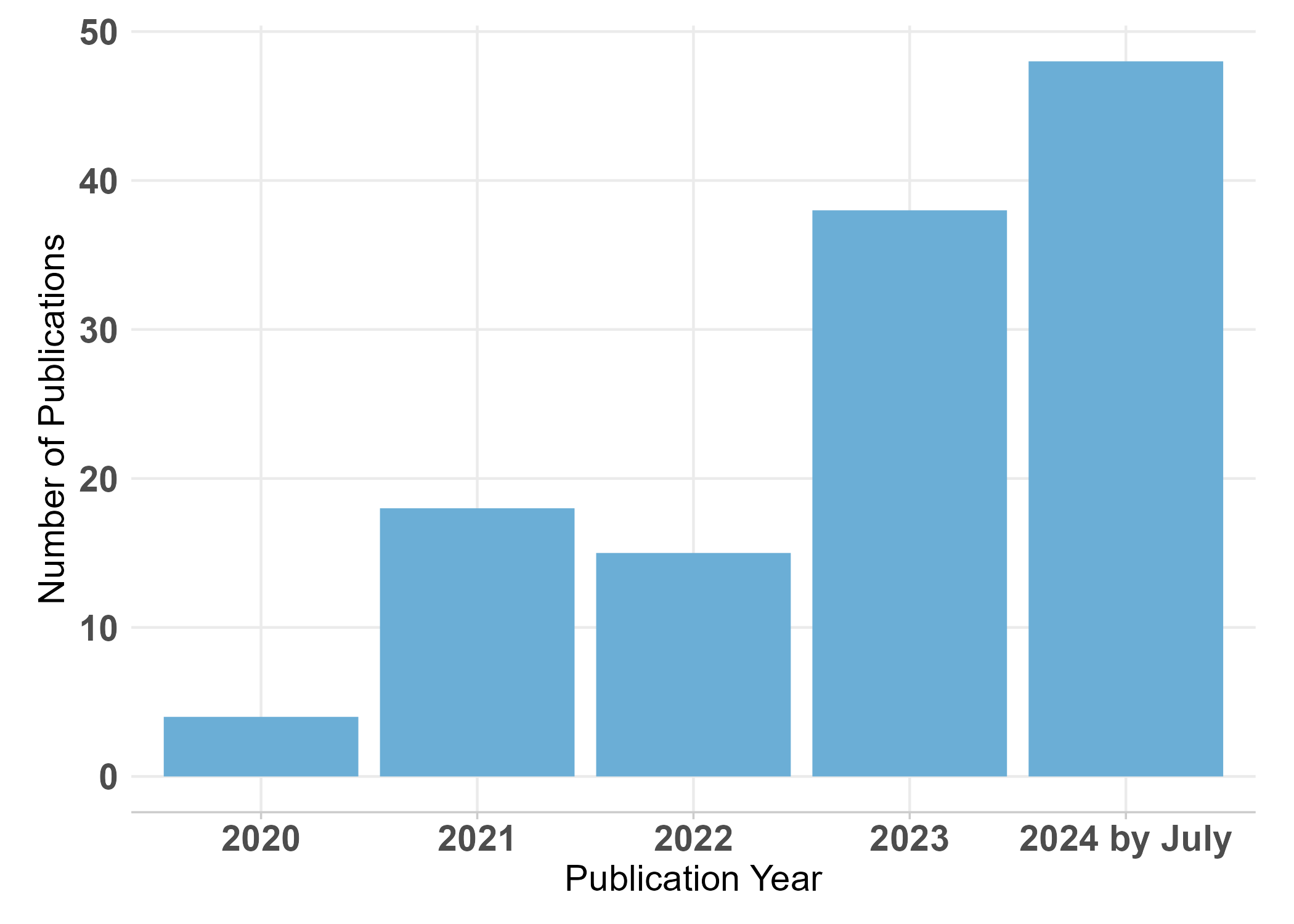}
    \caption{Number of Reviewed AI Literacy Publications Each Year}
    \Description{A graph that shows the number of Reviewed AI Literacy Publications Each Year}
    \label{pub_num}
  \end{minipage}
  \hfill
  \begin{minipage}[b]{0.49\textwidth}
    \includegraphics[width=\textwidth]{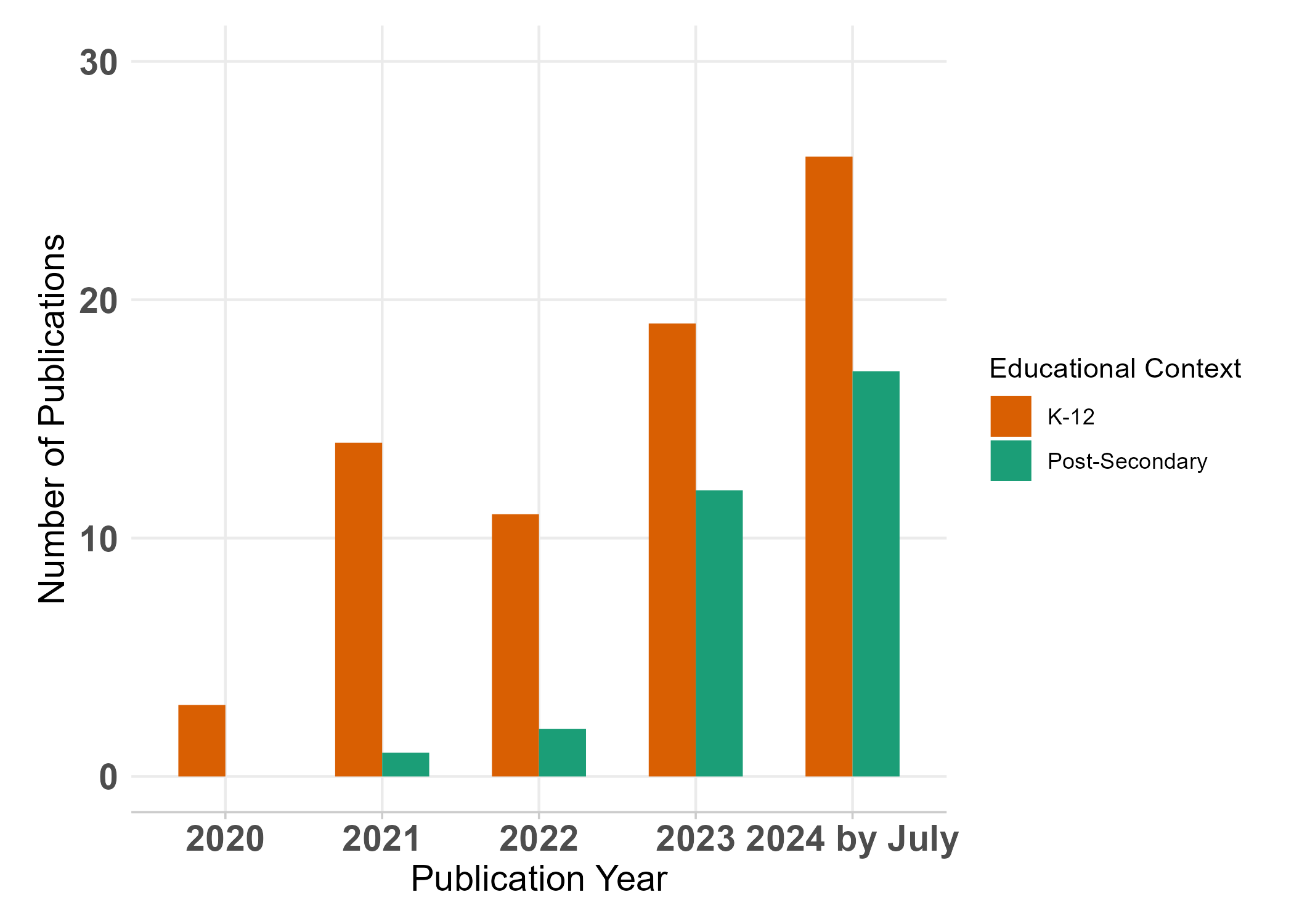}
    \caption{Educational Context of Reviewed Studies Each Year}
    \Description{A graph that shows the number of Reviewed AI Literacy Publications Each Year by their educational context, K-12 or post-secondary.}
    \label{K12_num}
  \end{minipage}
\end{figure}

Based on the number of reviewed studies published each year, several trends emerge. Note that the paper by Touretzky et al. is not included in the graphs, as it was included during purposive search due to its significant influence on the field, and not within in the main time range of this review. As shown in figure \ref{pub_num}, there were significant increases every year except between 2021 and 2022. Note that the literature search was conducted in July 2024, so the data for 2024 is not complete.  Despite that, the number of AI literacy publications in 2024 by July (48) already exceeded the total amount in 2023 (38). As shown in figure \ref{K12_num}, while there was a steady increase in the number of AI literacy studies in K-12 contexts, there was a much sharper increase in studies in post-secondary contexts. 

\begin{figure}[!htbp]
  \centering
  \hfill
  \begin{minipage}[b]{0.49\textwidth}
    \includegraphics[width=\textwidth]{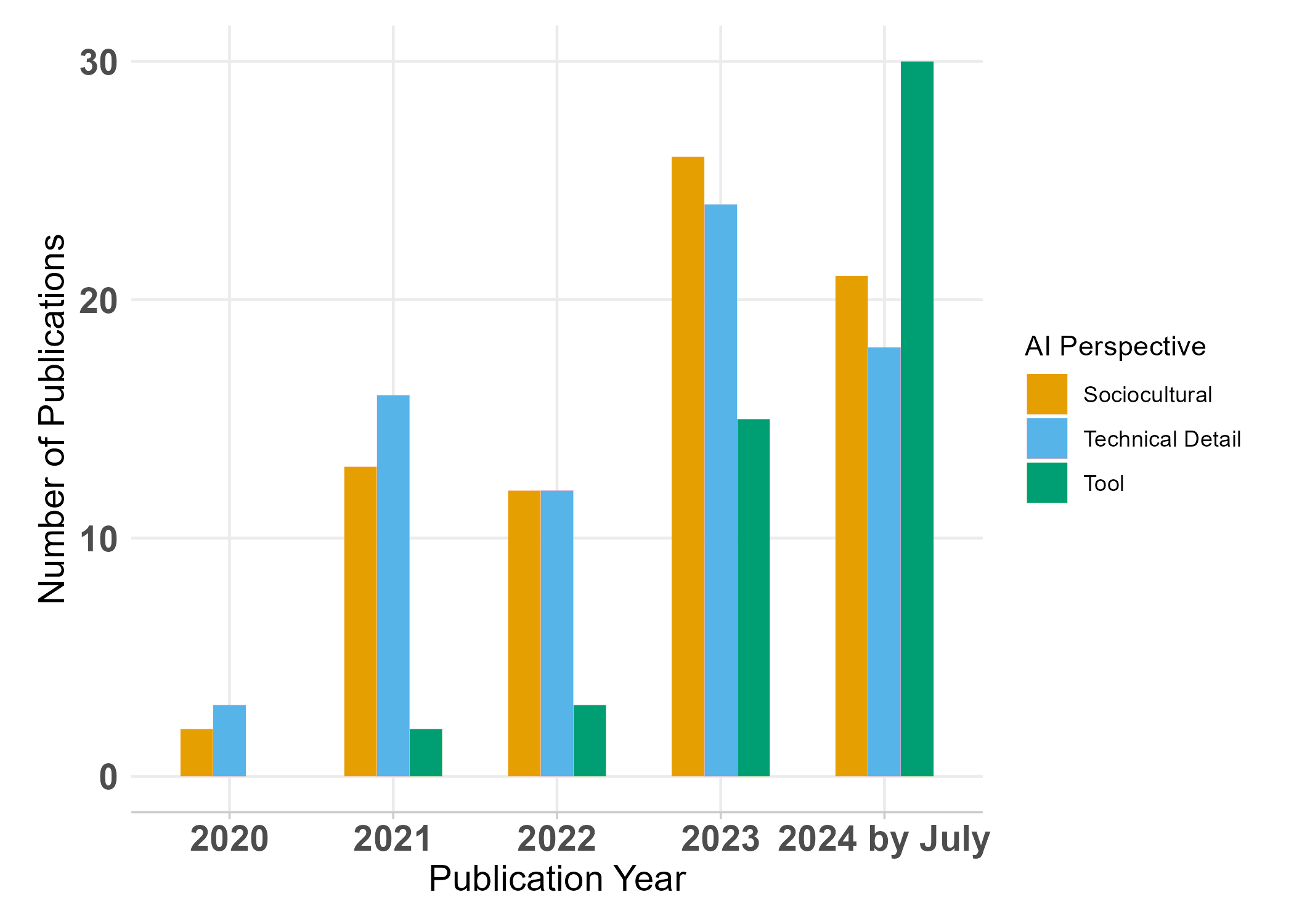}
    \caption{AI Perspectives of Reviewed Studies Each Year}
    \Description{A graph that shows the number of Reviewed AI Literacy Publications Each Year by their AI perspective, technical detail, tool, or sociocultural.}
    \label{AI_num}
  \end{minipage}
  \hfill
  \begin{minipage}[b]{0.49\textwidth}
    \includegraphics[width=\textwidth]{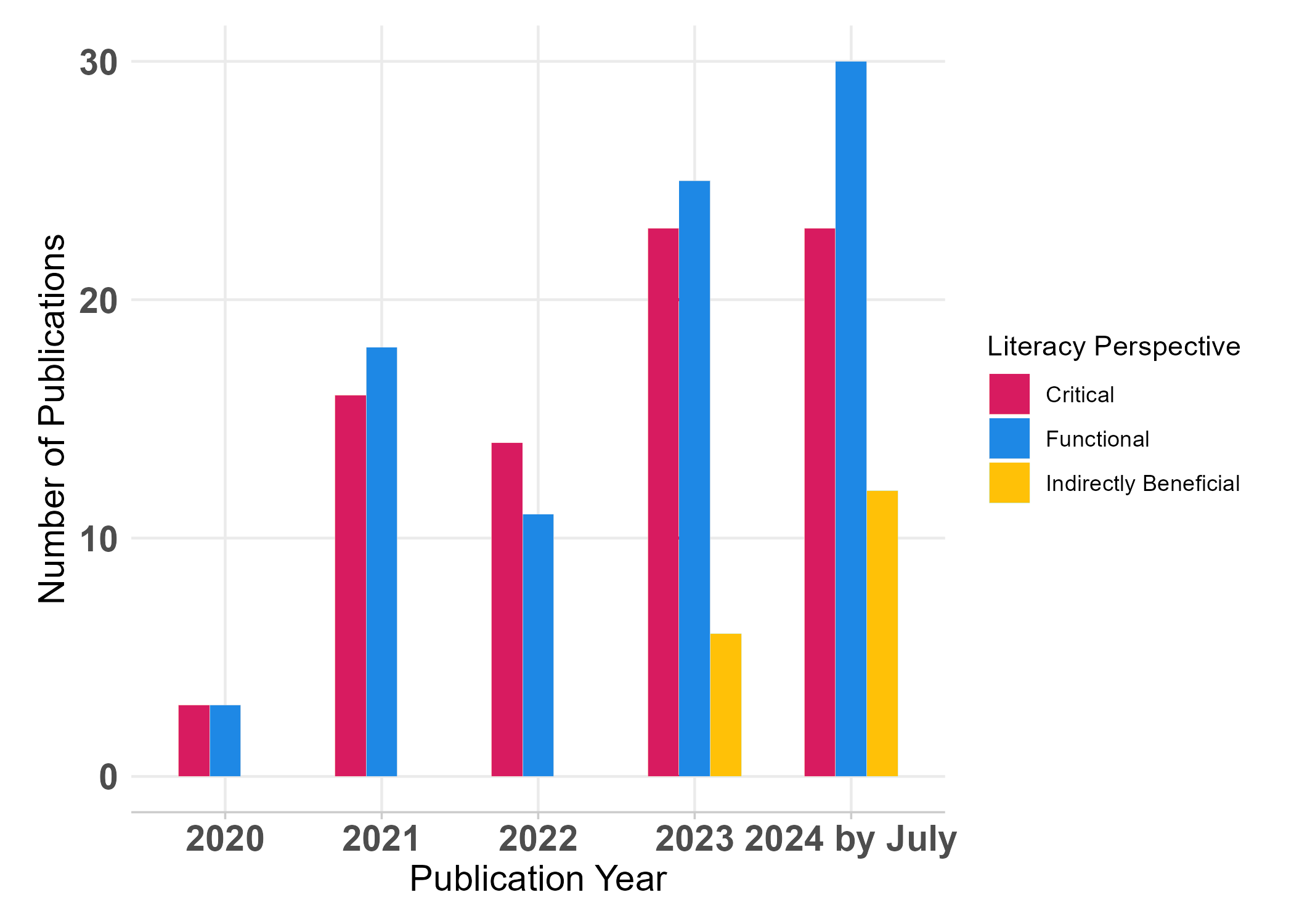}
    \caption{Literacy Perspectives of Reviewed Studies Each Year}
    \Description{A graph that shows the number of Reviewed AI Literacy Publications Each Year by their literacy perspective, functional, critical, or indirectly beneficial.}
    \label{lit_num}
  \end{minipage}
\end{figure}

As shown in figure \ref{AI_num}, while there were comparable numbers of studies that discuss AI by focusing on its technical details and its sociocultural influences each year, the number of studies that discuss AI as tools increased rapidly since 2023, dwarfing the other perspectives by July of 2024. As this review will later discuss, this trend was driven by the proliferation of AI literacy studies that specifically teach the use of generative AI tools. 

As for the literacy perspectives in figure \ref{lit_num}, there were comparable numbers of studies with a functional literacy perspective and critical literacy perspective over the time range of this review. However, starting in 2023, there was also the emergence of discussing AI literacy in terms of its indirect benefits, such as increasing students' interest in STEM, improving their computational thinking, or supporting their intrinsic motivation for studying CS. 

\begin{figure}[ht]
    \centering
    \includegraphics[width=0.7\linewidth]{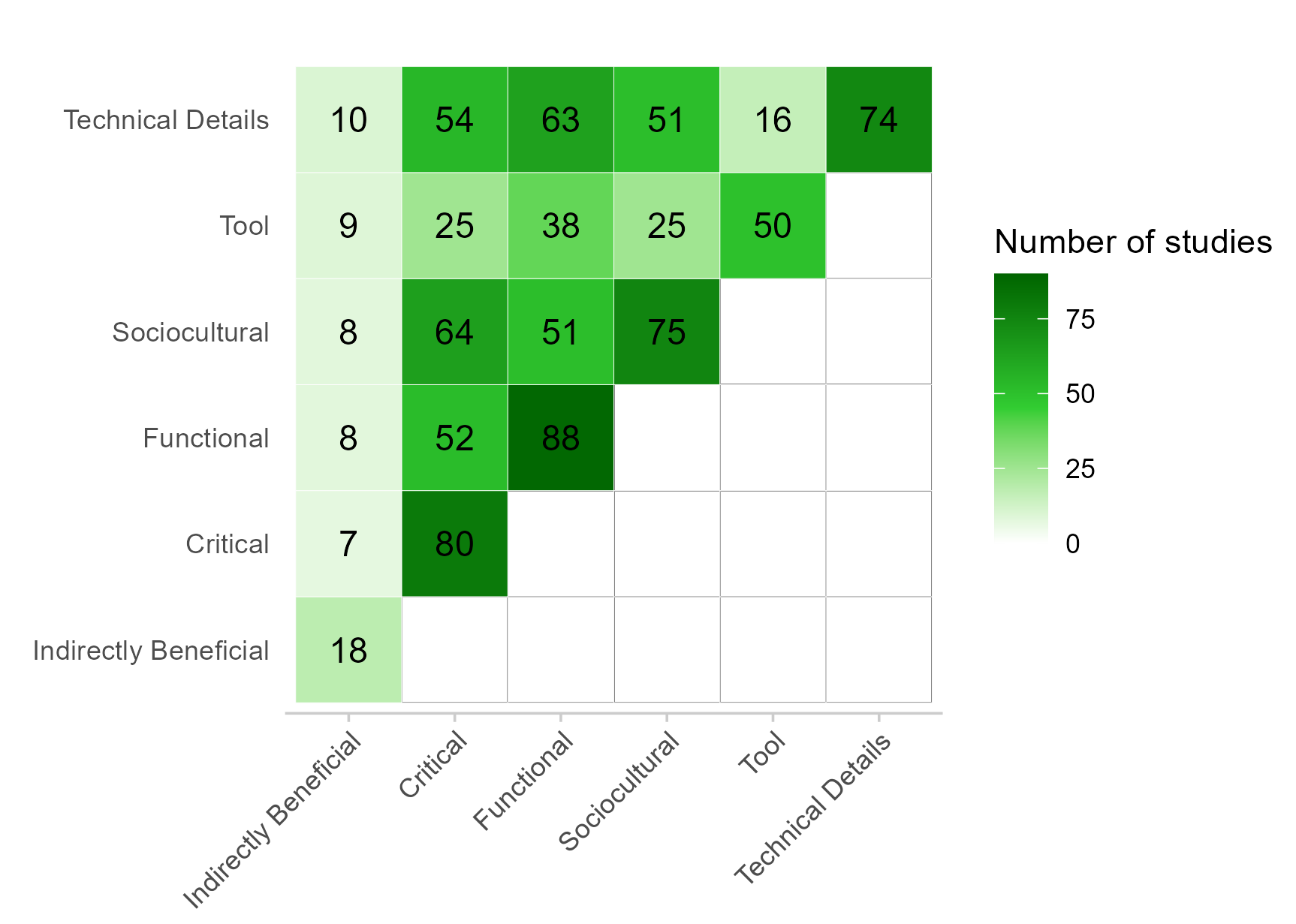}
    \caption{Number of Reviewed Studies that Shows Each Pair of AI or Literacy Perspectives (For example, 10 studies demonstrated both technical detail perspective of AI and indirectly beneficial perspective of literacy.)}
    \Description {A frequency matrix that shows the number of reviewed studies that shows each pair of AI or literacy perspectives.}
    \label{freq_mat}
\end{figure}

Figure \ref{freq_mat} shows the number of studies that include each combination of perspectives, which helps demonstrate how frequent different AI literacy approaches are studied. For instance, among the 88 studies that apply the functional literacy perspective, 63 of them also apply the AI technical detail perspective. This suggests that the relative popularity among reviewed studies to teach the technical fundamentals of AI, such as machine learning models, for functional stated goals, such as preparing students to become AI engineers. Other common approaches include pairing AI sociocultural perspective with critical literacy (e.g. teaching AI ethics to promote critical awareness regarding AI) and pairing AI tool perspective with functional literacy (e.g. teaching generative AI tool knowledge to ensure effective use). This graph also reveals notable research gaps. Relatively fewer studies combined an AI tool perspective with either a critical literacy or an AI sociocultural perspective, indicating a need for further research on teaching responsible and critical use of AI tools.

In the following detailed analysis, the reviewed studies are organized along several dimensions. To begin with, studies are sorted based on whether they involve generative AI. Note that this is not distinguished based on their publication time, as not all studies published after the public introduction of generative AI involved it. For instance, some studies published in 2024 continue to teach only machine learning curricula and do not mention generative AI (e.g. \citep{bilstrup_ml-machineorg_2024, lim_unplugged_2024}).
As such, this review focuses more on identifying established and emerging conceptualizations and approaches in the wake of generative AI, rather than identifying specific watershed moments that distinguish studies published before and after. 

Next, studies are sorted based on their educational context. For the purpose of this review, studies are divided broadly based on whether they focus on K-12 or post-secondary contexts. While there are variations in approaches to introduce AI literacy to high school students from those for elementary school students, there are more significant differences between K-12 and higher education that this review highlights. 

Lastly, studies within a specific educational context are organized based on their perspectives on AI and literacy. Each category will focus on commonalities of studies with its specific combination of AI literacy perspectives and highlight exemplar papers.

\subsection{Established AI Literacy Conceptualizations}

Several notable conceptualizations of AI literacy studies were established before the introduction of generative AI and have received continued interest since. These studies do not involve generative AI, even if they were published since 2023. There are numerous studies that introduced technical AI courses to K-12 settings with the stated objective of promoting AI literacy, even though similar courses in college CS major settings are almost never labeled as AI literacy (e.g. \citep{buxton_foundations_2024, baldoni_empowering_2021, zammit_road_2021}). In contrast, other studies specifically focus on raising awareness of AI ethics issues among K-12 students (e.g. \citep{dominguez_figaredo_responsible_2023, solyst_i_2023, schaper_robot_2020}). Another group of studies in K-12 settings use more comprehensive curricula that cover AI ethics and sociocultural implications in addition to AI technical details (e.g. \citep{lee_developing_2021, zhang_integrating_2023, williams_doodlebot_2024}). Lastly, there are a few AI literacy studies conducted in post-secondary contexts without involving generative AI, but they are relatively rare (e.g. \citep{kong_evaluating_2023, aler_tubella_how_2024}).

\subsubsection{Machine Learning for K-12}
Twelve studies (out of 124, 10\%) can be summarized as translating machine learning curricula that are traditionally taught in post-secondary settings to K-12 learners. Their contents are predominantly about machine learning principles and techniques, thus mostly framed from a technical detail perspective of AI. Two of these studies were published in 2020, constituting 50\% of the reviewed studies from that year. Four were from 2021 (22\% of that year's studies), one from 2022 (7\%), one from 2023 (3\%), and four from 2024 by July, when the literature search was conducted (8\%). The volume of research in this group remained low but stable. 

Six of the studies focuses on functional literacy \citep{buxton_foundations_2024, dwivedi_introducing_2021, quiloan_playgrid_2023, toivonen_co-designing_2020, bilstrup_ml-machineorg_2024, sanusi_learning_2024}. They extended the traditional perspective on AI as an academic branch of CS and adopted learning objectives that emphasize acquiring the skills and knowledge related to machine learning, such as implementing supervised learning algorithms to perform computer vision tasks \citep{buxton_foundations_2024}. Notably, all reviewed studies that use purely technical machine learning curricula were in K-12 settings, whereas similar curricula in higher education are rarely labeled as AI literacy. Three of these studies are about elementary school students \citep{dwivedi_introducing_2021, quiloan_playgrid_2023, toivonen_co-designing_2020}, two about middle school students \citep{bilstrup_ml-machineorg_2024, buxton_foundations_2024}, and one about a broader age range between 3-13 \citep{sanusi_learning_2024}.
These studies typically justify offering machine learning curricula to younger learners by emphasizing the importance of AI in the future. For instance, in the study by Buxton et al., they developed a curriculum that involved having twenty-nine 7-10 grade students train neural network-based image processing models that were then used to control autonomous robot cars \citep{buxton_foundations_2024}. 
When describing their study motivation, they cited how AI technology will transform future careers and K-12 educators need to prepare students for an “AI-driven society” \citep{buxton_foundations_2024}. 
The emphasis on adapting learners to potential future careers, along with the technical nature of the curriculum content, suggests a functional perspective on literacy. 

Six additional studies used similar curricula that focused almost entirely on machine learning, but mentioned their motivation to help learners become critical and empowered through their technical knowledge \citep{baldoni_empowering_2021, rodriguez-garcia_introducing_2020, voulgari_learn_2021, zammit_road_2021, sanusi_role_2022, zhang_developing_2024}.
For instance, in the most highly cited study in this group, Rodríguez-García et al. developed the LearningML platform that teaches exclusively ML algorithms, but stated their motivation was \textit{“to educate
critically thinking citizens able to understand technologies that
have a relevant impact on their lives”} \citep{rodriguez-garcia_introducing_2020}. Another study with Nigerian middle school learners implemented a technical machine learning curriculum, but emphasized bridging the gap between African AI education and the majority of AI literacy resources that are “Eastern and Western-centric” by considering “contextual and cultural values” to empower learners \citep{sanusi_role_2022}.

\subsubsection{AI Ethics in K-12}
Beyond treating AI as a technical field, many scholars highlighted the importance of AI ethics and recognized AI's sociocultural impact even before the rise of generative AI. While generative AI technologies represent the latest development, other machine learning applications have long been understood to have significant societal impact, such as facial recognition technologies gender and racial biases \citep{buolamwini_gender_2018}, political polarization from social media algorithms \citep{cho_search_2020}, and misuses of machine learning models in legal systems and policing \citep{angwin_machine_2016}. In response, another popular perspective in AI literacy research is viewing AI as sociotechnical systems and teaching learners about its sociocultural influences. Ten reviewed studies were in this category, with one (25\% of reviewed studies that year) in 2020, three (20\%) in 2022, four (11\%) in 2023, and two (4\%) in 2024 by July. Even though many of these studies were published after 2022, they do not involve generative AI.

From the sociocultural perspective of AI, researchers focus on educating learners about how AI becomes embedded in society and how they can participate as informed citizens. The majority of these studies adopt the critical literacy perspective. Their learning design approaches include having elementary school students write creative stories about solving real-world problems with AI \citep{ng_using_2022}, having students analyze automated decision-making systems and identify stakeholders to highlight accountability and fairness issues \citep{dominguez_figaredo_responsible_2023, xie_booklet-based_2022}, and inviting students to co-design AI technologies to express their visions of an AI-driven society and make them feel empowered \citep{lee_black_2022, solyst_i_2023}. The emphasis of these studies on identifying AI's social applications, stakeholders, and intentions behind their designs clearly highlight their sociocultural perspective on AI. There are also some effort at developing lecture-based curricula to teach AI ethics \citep{choi_effects_2024, kong_developing_2024, krakowski_authentic_2022}. Given these studies’ sociocultural focus, they usually spend little time on explaining the technical mechanisms of AI, instead often using open-ended discussion activities to engage students to think critically about how AI technologies interact with society. 
It is also worth noting that all these studies were conducted in K-12 settings, highlighting the gap in developing dedicated sociocultural AI or AI ethics learning activities for post-secondary settings. 

Even though a focus on the sociocultural perspective of AI pairs well with critical literacy, one reviewed study adopted the functional literacy perspective, despite focusing on AI as sociotechnical systems. Hammerschmidt and Posegga developed a AI literacy taxonomy that specifically focuses on sociocultural AI, but described their motivation as \textit{“sociocultural AI literacy plays a significant role in enabling employees to engage with AI effectively at work”} \citep{hammerschmidt_extending_2021}. 

\subsubsection{Comprehensive K-12 AI Curricula}
Without involving generative AI, the most prevalent AI literacy curriculum design was to first introduce AI’s technical aspects, then use them to explain its sociocultural effects. Thirty-four studies (out of 124, 27\%) were in this category. In contrast to the previous group adapting machine learning courses for K-12, many of these curricula expanded upon machine learning fundamentals with new learning activities that highlighted AI ethical issues and the sociocultural implications of AI technologies. Within the conceptual framework of this review, these studies have both technical detail and sociocultural perspectives of AI, and almost always included a critical perspective of literacy. The volume of reviewed studies in this category are high and stable over the years, even though its relative prevalence declined, with nine in 2021 (50\% of the reviewed studies that year), six in 2022 (40\%), eleven in 2023 (29\%), and seven in 2024 by July (15\%). 

There are a few prominent curricula that follow the combination of technical detail AI and a sociocultural AI perspective. Several studies developed curricula that follow the guidance from the AI4K12 initiative by the Advancement of Artificial Intelligence (AAAI) and the Computer Science Teacher Association (CSTA), which teaches the “five big ideas” of AI \citep{touretzky_envisioning_2019, yau_developing_2022, touretzky_machine_2023}. Three studies used another established curriculum is the Developing AI Literacy (DAILy) curriculum, where every unit covers technical knowledge (e.g. what is supervised learning), AI ethics (e.g. how supervised learning could cause algorithmic bias), AI awareness (e.g. identify technologies that use supervised learning), and AI career discussions \citep{lee_developing_2021, zhang_integrating_2023, zhang_effectiveness_2024}. Two research groups, prominently including Long et al., explored promoting AI literacy in informal learning settings, such as via museum exhibits where children can experience posture recognition technologies first-hand and family learning kits that teach supervised learning with an unplugged design \citep{lee_fostering_2023, long_co-designing_2021, long_role_2021, long_family_2022, long_ai_2023, long_fostering_2023}. As an example, in a highly cited study by Lee et al. published in 2021, the researchers held a workshop that implemented the DAILy curriculum for thirty-one middle school students, 87\% of which are from groups underrepresented in STEM \citep{lee_developing_2021}. The study conducted pre- and post-tests and found statistically significant increases in students' machine learning knowledge, interest in AI careers, and AI ethical concerns, such as latent biases \citep{lee_developing_2021}. Their stated objective was to help learners become \textit{“informed citizens and critical consumers of AI technology and to develop their foundational knowledge and skills to support future endeavors as AI-empowered workers”} \citep{lee_developing_2021}. This work serves as an example of how these interventions are designed with both functional and critical literacy in mind.

Numerous other studies in K-12 settings explored a wide range of approaches that combine technical and sociocultural AI perspectives. Some examples include design activities where students build robots or AI applications to learn both how they work and how they might be applied in real life \citep{van_brummelen_teaching_2021, williams_ai_2023, williams_doodlebot_2024, ng_fostering_2024-1, tanksley_were_2024}, programming activities that weave in ethical considerations (e.g. implementing a job application selection system using a decision tree that must be fair and accountable) \citep{druga_how_2021, kaspersen_high_2022, michaeli_data_2023, baldoni_does_2024}, developing a game-based learning experience such as a gamified ebook \citep{ng_fostering_2024} and a role-playing game of questioning a machine learning-based robot \citep{henry_teaching_2021} to surface machine learning principles and ethics, having learners exchange narrative stories about their experiences with AI online \citep{moore_teaching_2024}, dancing activities to teach posture recognition \citep{castro_ai_2022}, and traditional lectures and assessments that cover both technical AI knowledge and AI ethics \citep{khalid_familiarizing_2022, kong_evaluating_2023-1, ng_design_2024}. 
Even though their approaches are quite diverse, they all generally adhere to the goals of preparing students to work with AI and become informed and critical consumers of AI in the future \citep{eguchi_contextualizing_2021, kim_why_2021, steinbauer_differentiated_2021, wang_k-12_2023}. 
The diversity in approaches to designing AI literacy learning activities that promote both functional and critical literacy in K-12 contexts is encouraging, but also highlights a research gap in developing comparable AI literacy curricula for post-secondary settings.

\subsubsection{Post-Secondary AI Literacy without Generative AI}
AI literacy that does not involve generative AI in post-secondary contexts received relatively little research attention. As described in a 2022 scoping literature review by Laupichler et al., AI literacy in higher and adult education by then were almost entirely situated in specific professional settings, such as healthcare, or theorized AI applications in higher education \citep{laupichler_artificial_2022}. Neither of these meet the criteria of this review. Among studies included in this review, seven (out of 124, 6\%) were in post-secondary contexts and did not involve generative AI, with one in 2021 (6\% of that year's studies), one in 2022 (7\%), four in 2023 (11\%), and one in 2024 by July (2\%).

Further highlighting the gap in post-secondary AI literacy research, there was a distinctive lack of empirical learning intervention research in this category. The review identified three studies with empirical learning intervention components \citep{kong_evaluation_2021, kong_evaluating_2022, kong_evaluating_2023}. All of them were published by a group of researchers from the Education University of Hong Kong, who designed AI literacy modules for undergraduate non-CS major students. All involve comparatively short interventions, with between one and three modules that each last seven to nine hours, that serve as machine learning primers for non-CS major students \citep{kong_evaluation_2021, kong_evaluating_2022, kong_evaluating_2023}. 
The authors explicitly acknowledged that while the course content is strictly technical, lacking ethics components, their goal is to empower non-CS major students regarding AI technologies by demystifying their technical aspects \citep{kong_evaluation_2021, kong_evaluating_2022, kong_evaluating_2023}. As such, these studies present an example of combining the technical detail perspective of AI with critical literacy.

The other studies in this group did not include empirical components. Aler Tubella et al. conducted a literature review and interviewed eleven experts to develop a post-secondary AI literacy framework, with the stated goal of ensuring \textit{"the next generation of young people can contribute to an ethical, safe and cutting-edge AI made in Europe"} \citep{aler_tubella_how_2024}. Interestingly, this study discussed AI literacy completely in terms of ethics, governance, and policy making, while never touching on the technical aspects of AI or generative AI \citep{aler_tubella_how_2024}. It is thus an outlying example of applying only the sociocultural perspective of AI and critical perspective of literacy.
The remaining three studies are about developing AI literacy assessments instead of interventions, which all include test items that assess both students' machine learning and AI ethics knowledge \citep{hornberger_what_2023, laupichler_development_2023, laupichler_evaluating_2023}.

\subsection{Emerging AI Literacy Conceptualizations}
Due to the significantly improved accessibility of AI technology enabled by natural language interfaces, learners can directly interact with generative AI tools. This caused both concerns and excitement among researchers and practitioners. Since the rise of generative AI, there are several new kinds of AI literacy that emerged that often directly address generative AI tools: effective usage of generative AI tools for post-secondary students (e.g. \citep{chiu_future_2024, shibani_untangling_2024, denny_prompt_2024, sengewald_teaching_2024}), incorporating AI tools into K-12 AI literacy curricula (e.g. \citep{kajiwara_ai_2024, solyst_childrens_2024, relmasira_fostering_2023}), and increased mentions of the indirectly beneficial perspective of AI literacy (e.g. \citep{su_span_2024, kim_change_2023, lim_unplugged_2024}). It is also worth mentioning that generative AI tools can have user policies that restrict use based on age. For example, ChatGPT user policy stipulates that children under 13 should not use the tool and children between 13 and 18 need parental or guardian consent before using. These policies might influence the feasibility of K-12 AI literacy research on generative AI.

\subsubsection{Teaching Generative AI Tool Use}
Twenty-two reviewed AI literacy studies (out of 124, 18\%) adopt only the tool perspective of AI, and several discuss teaching ChatGPT use specifically (e.g. \citep{hu_explicitly_2023}). Three of these studies were published in 2023 (8\% of the reviewed studies from that year), and nineteen (40\%) were published in 2024 by July. Notably, sixteen of these studies were about post-secondary learners, already more than the total of post-secondary AI literacy studies that do not involve generative AI. 

Among the studies that exclusively view AI as tools and view literacy as functional, almost all are concerned with teaching or assessing students on effective use of generative AI tools. This framing of AI literacy is best captured by a study that proposes an AI literacy definition in terms of skills of using generative AI tools: \textit{“future higher education should be transformed to train students to be future-ready for employment in a society powered by GenAI”} \citep{chiu_future_2024}. Especially given LLMs’ ability to generate computer programs, CS educators have concerns over students potentially abusing LLM-based tools \citep{lau_ban_2023}.
To address this, seven studies isolated prompt engineering as a specific competency that should be promoted to further AI literacy, and have devised many ways to let students practice it \citep{mabrito_artificial_2024, knoth_ai_2024, walter_embracing_2024, haugsbaken_new_2024, denny_prompt_2024, varuvel_dennison_consumers_2024, zamfirescu-pereira_why_2023}. 
For instance, Haugsbaken and Hagelia theorized 
that \textit{“as students and teachers engage with AI language interpreters on a daily basis”}, prompt engineering should be \textit{"considered to be part of a future AI literacy"} \citep{haugsbaken_new_2024}. 
In practice, researchers have developed practice problems that task students to write prompts that can let ChatGPT generate a program with the desired output \citep{denny_prompt_2024}, and an assessment of students’ prompt engineering skills that identified common ineffective prompting strategies by students (e.g. treating prompting as a human-to-human instruction) \citep{zamfirescu-pereira_why_2023}. 

Alternatively, some scholars investigated directly teaching CS major students proper usage of ChatGPT as a form of AI literacy \citep{fernandez_cs1_2024, hu_explicitly_2023, husain_potentials_2024}. For example, Hu et al. designed an undergraduate CS1 assignment that introduced students to ChatGPT, provided guidance on how to use it effectively and ethically, and required self-evaluations on whether they followed the guidance \citep{hu_explicitly_2023}. They found that an explicit introduction to ChatGPT helped engage students that had a low attendance rate \citep{hu_explicitly_2023}. ChatGPT is frequently mentioned among these studies as the main tool that students should become adept with, but some studies explored skillful use of generative AI tools in general \citep{delcker_first-year_2024, varuvel_dennison_consumers_2024}.

Along with the recognition of generative AI tools’ potentials, researchers and practitioners also express concerns over them, such as over-reliance and latent biases in LLMs among others \citep{lau_ban_2023}. 
Consequently, some AI literacy researchers adopted a critical literacy perspective regarding generative AI tools. While researchers have quickly devised various interventions to improve students’ functional literacy about generative AI tools, it is less clear how to promote their critical literacy, as several studies surveyed or interviewed students about their AI tool usage behavior to determine what concerns should be addressed \citep{belghith_testing_2024, celik_exploring_2023, chen_exploring_2024}. For instance, Chen et al. surveyed undergraduate students about their ChatGPT use and concluded that while students are aware that they should be critical of generative AI output, they still expressed the need for \textit{“explicit guidance from course syllabi and university policies regarding generative AI’s ethical and appropriate use”} \citep{chen_exploring_2024}. Two studies designed AI tool critical literacy interventions, including a card-based learning activity \citep{wang_card-based_2024} and a series of AI tool ethics modules \citep{ariyarathna_teaching_2024}, however both were in early stages and were not deployed in educational settings. 

\subsubsection{Incorporate Generative AI Tools in AI Literacy Curricula}

While the prevailing AI literacy curriculum before the rise of generative AI combined the technical detail perspective and the sociocultural perspective of AI, the availability of generative tools provides an opportunity to include more hands-on activities. As such, twenty-five studies (out of 124, 20\%) adopted multiple perspectives of AI that included the tool perspective. Notably, ten of them were in post-secondary contexts, in contrast to studies in K-12 contexts monopolizing this type of AI literacy curricula. One study in this group was published in 2021 (6\% of that year's studies), two in 2022 (13\%), eleven in 2023 (29\%), and eleven in 2024 by July (23\%). Note that while the studies published before 2023 used tools that did not involve generative AI, they are analyzed in this section to illustrate the rising trend.

Multiple studies explored teaching AI ethics concerns with generative AI tools to K-12 learners \citep{su_artificial_2023, su_artificial_2022, yang_artificial_2024, su_span_2024, marienko_artificial_2024, okolo_beyond_2024, kajiwara_ai_2024, solyst_childrens_2024, ali_children_2021}. 
Starting with Kindergarten settings, a research group from the University of Hong Kong focused their research on early childhood AI literacy, which involved having students interact with social robots and explore their inner workings \citep{su_artificial_2023, su_artificial_2022, yang_artificial_2024, su_span_2024}. 
In middle school settings, five studies had researchers demonstrating AI tools to students without having them interface with the tool directly \citep{marienko_artificial_2024, okolo_beyond_2024, kajiwara_ai_2024, solyst_childrens_2024, ali_children_2021}. 
For instance, to teach learners about ChatGPT's hallucination and latent biases and promote their critical literacy, researchers would show examples of ChatGPT making obvious mistakes, then guide students to consider its causes and societal implications \citep{okolo_beyond_2024, kajiwara_ai_2024, solyst_childrens_2024}. 
Two studies in K-12 settings involved some hands-on interaction with AI tools, including ChatGPT \citep{vo_ai_2023} and DALL-E \citep{relmasira_fostering_2023}. These activities have students explore these tools under instructor supervision and reflect upon their sociocultural implications, such as asking students what interacting with DALL-E taught them about targeted misinformation \citep{relmasira_fostering_2023}. Both hypothesized that introducing AI tools early could help students learn “responsible and ethical use”, indicating their functional and critical literacy perspectives \citep{relmasira_fostering_2023, vo_ai_2023}.

Studies in post-secondary settings faced less restrictions in having students use AI tools. Combined with heightened interest in generative AI in higher education institutions, there are now larger-scale efforts at developing college curriculum that teaches students both effective tool usage and critical understanding of the effect on the future labor market \citep{brew_towards_2023, hemment_ai_2023, salhab_ai_2024, sengsri_artificial_2024, southworth_developing_2023, tenorio_teaching_2023, flechtner_making_2024}. These efforts go beyond targeting CS major students,
such as a group of University of Florida researchers arguing for \textit{“infusing AI across the curriculum and developing opportunities for student engagement within identified areas of AI literacy regardless of student discipline”} \citep{southworth_developing_2023}. While it is expected that post-secondary AI literacy initiatives often have the functional literacy perspective, over half of them also adopted the critical literacy perspective. For instance, Gupta et al. theorized how metaphors used to describe ChatGPT, such as an assistant or a parrot, can be used to promote critical reflections among students \citep{gupta_assistant_2024}. 
AI literacy studies in post-secondary settings still involve empirical evaluations of learning interventions less frequently than studies in K-12. The remaining post-secondary studies developed AI literacy modules \citep{shi_build_2024} or assessments \citep{biagini_developing_2024, faust_assessment_2023, hwang_development_2023} that cover all perspectives of AI and both functional and critical literacy.

\subsubsection{Indirect Benefits of AI Literacy}
Lastly, there is a curious emergence of AI literacy research that was motivated by reasons other than promoting functional or critical literacy. They cite various other benefits, such as increased interest in STEM, improved computational thinking, or having a more positive attitude towards AI. Six were published in 2023 (16\% of reviewed studies published that year) and twelve in 2024 by July (25\%). 
Nine of these studies did not feature the perspective of indirectly beneficial literacy prominently and are already discussed previously \citep{baldoni_does_2024, benvenuti_artificial_2023, moore_teaching_2024, ariyarathna_teaching_2024, ng_design_2024, ng_fostering_2024-1, walter_embracing_2024, kajiwara_ai_2024}. This section will discuss the rest.


Five papers that taught machine learning curricula to K-12 students justified it by citing benefits that are not directly relevant to AI, namely better computational thinking efficacy \citep{lin_modeling_2024}, motivation and self-efficacy \citep{kit_ng_artificial_2023}, increased interest in STEM \citep{lim_unplugged_2024, vandenberg_promoting_2022}, more positive attitude towards AI technologies \citep{kim_change_2023}, and creative thinking \citep{lim_unplugged_2024}. For example, in a study that teaches the underlying machine learning principles in facial recognition technology (FRT) to middle school students, Lim et al. argued that \textit{“AI education has many potential benefits, such as fostering creative thinking and motivation in K-12 students”} \citep{lim_unplugged_2024}. Notably, this study did not discuss the ethical implications, such as possible racial biases, in FRTs, instead it exclusively discussed supervised learning mechanisms. These indirect benefits of learning AI can neither be characterized as functional literacy or critical literacy, and are thus classified as indirect benefits. 

Lastly, four studies that taught learners how to use AI tools did not appeal to either functional or critical literacy, but instead cited other benefits. One involved a workshop where children of age 10 to 12 interacted with an LLM-based music generation tool to teach music and foster interest in STEAM (arts in addition to STEM subjects) \citep{parada-cabaleiro_techvilles_2024}. Quoting a European Commission report, the author emphasized promoting AI literacy \textit{“for personal fulfillment, a healthy and sustainable lifestyle, employability, active citizenship and social inclusion”} \citep{parada-cabaleiro_techvilles_2024}. 
The other three related studies involved having kindergarten learners interact with social robots to increase their interest in AI \citep{su_artificial_2023, su_artificial_2023-1, yang_artificial_2024}. The researchers of these studies acknowledged that \textit{“young children at this age hardly know and understand AI”} given the mean age of participants was four, and the AI literacy interventions were developed to instill favorable attitudes to AI and possibly improve children's digital literacy skills in preparation for elementary school \citep{su_artificial_2023, su_artificial_2023-1, su_span_2024}.

\section{Discussion}
\subsection{\textit{RQ1: How do researchers conceptualize and approach AI literacy in K-12 and higher education?}}

To address this research question, this study reviewed 124 studies on AI literacy published since 2020 in K-12 or post-secondary settings and constructed a conceptual framework to capture the different conceptualizations of AI literacy, shown in figure \ref{framework}. This conceptual framework organizes the wide range of studies that promote AI literacy based on their perspectives on AI and literacy. Regarding AI, this work is informed by frameworks, including the Dagstuhl Triangle and learner roles around AI \citep{brinda_dagstuhl-erklarung_2016, faruqe_competency_2022, schuller_better_2023}. It categorizes studies based on whether their AI literacy conceptualizes AI as its technical details (e.g. machine learning fundamentals), as tools (e.g. ChatGPT), or as sociotechnical systems with sociocultural impact (e.g. facial recognition technology and its privacy and surveillance implications). Regarding literacy, this work is informed by literacy theory and digital literacy frameworks \citep{scribner_literacy_1984, selber_multiliteracies_2004} and includes three perspectives of literacy: functional literacy (e.g. understand and implement machine learning models), critical literacy (e.g. critically evaluate AI products and use them responsibly), and inherently beneficial literacy (e.g. become more interested in STEM). 

The significant differences between perspectives can help explain the difficulty researchers have had with creating a clear definition of AI literacy. It highlights the inherent ambiguity in the framing of AI literacy, as the various perspectives on AI and literacy are all valid and apply to different learning contexts and learning objectives. This framework does not contradict existing frameworks, nor does it seek to simplify them. Instead, existing taxonomies of AI competencies are essentially more granular lists of skills under categories in this framework, such as specific skills students should master to achieve functional literacy with technical AI \citep{shiri_artificial_2024}. Instead, the framework demonstrates the need to use more specific terms when discussing skills and knowledge surrounding AI, as the umbrella term of “AI literacy” encompasses a broad range of skills that need to be implemented very differently in curricula in practice. 
Describing conceptualizations of AI and literacy with greater specificity can help researchers and practitioners communicate their objectives and approaches more clearly. 


The review also identified gaps in current research. The AI tool perspective is relatively infrequently combined with the critical literacy perspective or the sociocultural AI perspective, even though these combinations are crucial for responsible and safe use of AI tools. For instance, learning to critically evaluate generative AI tools and their outputs can help students recognize issues such as the creation and spread of disinformation using generative AI \citep{solyst_childrens_2024}. Another research gap is in combining the technical details AI perspective with the AI tool perspective. While there are relatively few studies of these perspectives, there are empirical studies that suggest teaching students about the technical aspects of AI can help them better use AI tools \citep{relmasira_fostering_2023, mabrito_artificial_2024}. 

\subsection{\textit{RQ2: What new trends have emerged in AI literacy research in K-12 and higher education in the wake of generative AI?}}

By examining AI literacy studies that were published around the public launch of generative AI technologies, this review surfaced emerging trends in research interests and shifts in conceptualization of AI literacy. Among studies that do not involve generative AI, there has been sustained effort in promoting AI literacy in K-12. There are several approaches to teaching AI literacy in K-12 before generative AI. These include adapting machine learning curricula that are traditionally taught at college-level to K-12 learners, teaching AI ethics and the social impact of machine learning technologies to K-12 (e.g. latent racial and gender biases of facial recognition technology), and comprehensive AI literacy curricula that combine both technical details and ethics of AI. 

There are relatively few post-secondary AI literacy studies that do not involve generative AI. Several of them developed curricula or assessments, without deploying and testing them empirically. The only empirical AI literacy research in post-secondary contexts that does not include generative AI was about teaching machine learning concepts to non-CS major students \citep{kong_evaluation_2021, kong_evaluating_2022, kong_evaluating_2023}. This pattern is likely due to the fact that AI courses in post-secondary contexts were almost never labeled as AI literacy prior to the rise of generative AI.

Since the rise of generative AI, there are three major shifts in AI literacy conceptualization among the reviewed studies. The first of which is a shift toward teaching generative AI tool use in post-secondary contexts. 
Due to the restrictions surrounding having learners under the age of 18 use generative AI tools, AI tools are more accessible to post-secondary students. Consequently, 
researchers in post-secondary contexts expressed greater interests in AI applications and concerns over potential AI tool misuse \citep{lau_ban_2023, prather_robots_2023}. In response, a significant number of studies that specifically tackle post-secondary students' generative AI tool use were published since 2023. For instance, seven studies focus on improving students' prompt engineering skills specifically. However, there is still a gap in empirical studies in post-secondary contexts, as many studies developed tools, assessments, and curricula without empirically evaluating them. 

Another trend is in K-12 AI literacy curricula incorporating AI tool perspectives. While the restrictions of AI tool use by younger learners still applies, researchers and practitioners have explored ways to familiarize K-12 learners with AI tools, such as by having them interact with more limited social robots or demonstrating the uses and issues with generative AI tools without having students interface with them \citep{su_span_2024, solyst_childrens_2024}. These efforts lead to a more comprehensive AI literacy curricula that includes more concrete and hands-on learning activities, supporting both functional and critical literacies of learners. There remains a research gap in translating these curricula to post-secondary contexts.

Lastly, coinciding with the rise of generative AI, there is an emerging trend of discussing AI literacy in terms of its indirect benefits, such as increased interest in STEM or improved computational thinking. While not all AI literacy studies with this perspective of literacy involve generative AI, all these studies were published following the public launch of generative AI. This correlation could be attributed to increased interest surrounding AI in general since the rise of generative AI, which in turn attracted researchers and practitioners with broader research interests to study the potential outcomes of teaching AI. 

To summarize, AI literacy research before generative AI was focused on K-12, placing relatively equal emphasis on functional and critical literacies, and teaching both technical details and sociocultural influences of AI. Since generative AI became popular, the conceptualization of AI literacy shifted rapidly toward more research in post-secondary settings on AI tool use, often with a functional literacy perspective. There is also an increase in studies that discuss AI literacy based on its indirect benefits. There are research gaps in promoting critical literacy surrounding AI tools, and conducting more empirical AI literacy research in post-secondary contexts in general. There also remains a general lack of empirical research on AI literacy interventions in post-secondary settings, which suggests a need for greater research attention and resource to promoting AI literacy among higher education students.

\section{Limitations}
This review has several limitations. It applies an integrative review methodology to generate a framework of AI literacy, but does not seek to address specific research questions within the field of AI literacy. It made no attempt at evaluating the effectiveness of the learning interventions within the studies, because the learning interventions are not directly comparable given their differences in research questions and methodologies. The relative frequency of AI literacy conceptualizations should not be interpreted as an indication of their effectiveness. This review also does not include AI literacy within several education settings that each have a significant amount of relevant studies, including pre-service teacher AI literacy, adult AI literacy, and AI literacy in healthcare, nursing, and business contexts. It requires future work to verify if the conceptual framework can be generalized into those contexts.

\section{Conclusion}
Neither AI nor literacy are simple concepts. AI literacy needs to be understood as a set of connected but distinct skills and knowledge, where each type is crucial in different contexts. An AI engineer working on LLMs would need deep functional literacy on AI technical details. In contrast, citizens voting on AI regulation and consumers deciding on which AI products to use need critical literacy on AI as tools and AI’s sociocultural impact. Informed by prior theoretical work, this review of 124 studies on AI literacy in K-12 and post-secondary contexts constructs a conceptual framework of three perspectives on AI: technical detail, tool, and sociocultural, and three perspectives on literacy: functional, critical, and indirectly beneficial. 

This review examined trends in AI literacy in the wake of generative AI, and found major shifts in AI literacy research and conceptualization. AI literacy before generative AI was primarily studied in K-12 contexts, where K-12 learners were introduced to both the technical details of AI and AI ethics, so that they could develop both functional and critical AI literacy. In the wake of generative AI, there is significantly more research conducted in post-secondary contexts, usually with an AI tool perspective. They emphasize teaching effective use of generative AI tools, such as by teaching prompt engineering. K-12 AI literacy research has similarly adapted by incorporating more AI tool elements. There is also an emergence of studies that discuss the indirect benefits of AI literacy, such as increased interest in STEM, since 2023. 

This work can help researchers and practitioners more clearly conceptualize and communicate their objectives and approach to AI literacy, thus encouraging more effective discussion and knowledge generation surrounding this crucial topic. The ambiguity in the generic term of “AI literacy”, as shown by the drastically different interventions that share the same label, suggests a need for using more specific terms based on different conceptualizations of AI and literacy. This review also identifies potential gaps in AI literacy research, such as promoting critical literacy among students when interacting with AI tools. If AI tools become as prevalent in the future as many researchers and educators anticipate, these research gaps represent valuable future research directions. 





\bibliographystyle{ACM-Reference-Format}
\bibliography{sample-authordraft}


\end{document}